\documentstyle[aps,preprint]{revtex} 
\tightenlines 
\begin{document}

\title{Rain, power laws, and advection} 
 
\author{Ronald Dickman$^\dagger$ 
} 
\address{ 
Departamento de F\'{\i}sica, ICEx, 
Universidade Federal de Minas Gerais,\\
Caixa Postal 702, 
30161-970 Belo Horizonte, MG, Brazil} 
\date{\today} 
 
\maketitle 
\begin{abstract} 
Localized rain events have been found to follow power-law
size and duration distributions over several decades, 
suggesting parallels between precipitation and seismic
activity [O. Peters et al., PRL {\bf 88}, 018701 (2002)].
Similar power laws are generated by treating 
rain as a passive tracer undergoing advection in a velocity field
generated by a two-dimensional system of point vortices. 
\vspace{2em}  

\vspace{2em}

\noindent $^\dagger${\small Email address: dickman@fisica.ufmg.br}

\end{abstract} 

 \newpage

As natural phenomena are probed on ever finer length and time
scales, surprising scale-invariant properties are brought to light.
A notable instance is the recent discovery that 
distributions of
rainy and dry intervals, and the size of rain events,
follow power laws \cite{peters,christensen}.
Analyzing Doppler radar data, Peters, Hertlein and Christensen
found that the distribution of local rain-event sizes decays as a
power law over at least three decades. Durations 
of rain-free intervals (`droughts') are also power-law distributed over the 
range of several minutes to about a week, with a significant 
perturbation reflecting diurnal variation.

The similarity between these observations and scaling laws in
seismic activity (the Gutenberg-Richter law for earthquake sizes, and
the Omori law for waiting times) suggests a parallel between
atmospheric precipitation and relaxation of the Earth's crust at
stressed tectonic-plate boundaries \cite{christensen}.  In the
latter context, cooperative relaxation due to elastic interactions
and nonlinear friction is captured by block-spring models 
\cite{burridge,carlson} or, in much-reduced fashion, by
sandpile models \cite{btw}.  
But if certain aspects of rain distributions resemble those of avalanches in
sandpile-like models, the underlying physics remains obscure.
There is no obvious reason for the formation or precipitation 
of one raindrop to provoke similar events nearby.  
Given the attendant release of latent heat,
one would instead expect a self-limiting tendency in condensation.
Thus it appears more promising to seek the explanation for
power-law distributions in atmospheric dynamics.

Atmospheric motion involves turbulence,
particularly in the vicinity of storms; various aspects
of turbulent flow follow power laws 
over many orders of magnitude \cite{mccomb,frisch}.
Even in the absence of fully developed turbulence, 
unsteady flow will stretch and fold any
initially compact region, leading to a highly convoluted,
nonuniform density of suspended particles or droplets 
\cite{tritton,elperin} via chaotic advection \cite{aref84}.
In light of these observations, it is interesting to develop a
model in which rain is an {\it ideal passive tracer}, following
the local fluid velocity {\bf u}({\bf x}$(t),t$) 
\cite{mccomb12,falkovich}. 
({\bf x}$(t)$ denotes the instantaneous position of the tracer.)
This Letter aims to show that such a model is capable of producing 
power-law-distributed event sizes and durations.

I adopt a simplified model susceptible to
numerical simulation.  A convenient alternative
to numerical integration of the Navier-Stokes equation is
afforded by vortex dynamics \cite{mccomb}.  In the two-dimensional
case studied here, a system of $N_V$ point vortices \cite{sommerfeld}, 
each moving in the velocity field generated by all vortices except itself,
is a particularly attractive method for simulating incompressible, 
inviscid flow. 
Point-vortex systems have been used
for some time in studies of two-dimensional 
turbulence \cite{mccomb,aref,kraichnan}. 

The simulation cell is a unit square with periodic boundaries.   
The velocity of vortex $i$ is given by
\begin{equation}
{\bf v}_i = \sum_{j \neq i} \frac{K_j}{2 \pi r_{ij}^2} 
\hat{\bf k} \times {\bf r}_{ij} ,
\label{vorvel}
\end{equation}
where $K_j$ represents the circulation of vortex $j$ 
(equal numbers of clockwise and anticlockwise vortices are used),
and {\bf r}$_{ij} = {\bf r}_i - {\bf r}_j$, under period boundaries, 
using the nearest-image criterion.  The velocity {\bf u}({\bf x},$t$)
at an arbitrary point {\bf x} in the plane (not occupied by a vortex)
is given by a similar sum including contributions from all vortices.
The number of vortices $N_V$ ranges from 10 to 126.

I study several types of vortex-strength distributions.
The simplest assigns all vortices the
same strength $|K|$.  
Other studies employ a hierarchical vortex
distribution, defined as follows.  The zeroth ``generation" consists
of a pair of vortices with $K = \pm K_0$.  Subsequent generations,
$n = 1,...,g$ have $2^{n+1}$ vortices, with circulation $|K| = K_0/\alpha^n$. 
I study $\alpha$ = 2, 3, and 4,  
using $g\!+\!1 = 5$ or 6 generations. 
The purpose of 
the hierarchical distribution is to provide structure on a variety of
length scales, without trying to reproduce any specific energy spectrum 
$\epsilon (k)$.  
The vortices are assigned random initial positions, 
but their subsequent evolution is deterministic.  
Being point objects, they possess no intrinsic length scale.
(Note however that in the presence of other vortices, the `sphere of
influence' of vortex $i$ is proportional to $K_i$.)
A characteristic length scale 
is the mean separation $\sim 1/\sqrt{N_V}$ between vortices.  
The vortex system defines a mean speed 
$u = \langle |{\bf u}({\bf x},t)| \rangle \propto K_0 \sqrt{N_V}$;
an important time scale is $\tau_C \sim 1/u$, the typical time for a fluid
particle to traverse the system.

A large number of
tracers, $N_p = 5\;000$ or $10\;000$, are thrown 
at random into a small region (of linear dimension 0.05) to simulate
a localized condensation event.  
(Alternatively, the tracer-laden region may be interpretated
as a parcel of atmosphere of high humidity, destined to
generate precipitation.)
`Rain' corresponds to the presence of one or more
tracers in a very small predefined region or `weather station',
of linear dimension 0.01.  (To improve statistics, I use 40 - 50 such
stations; each is randomly assigned a fixed position.)  At each
step of the integration ($\Delta t = 5 \times 10^{-6} - 10^{-4}$), the number
of particles $n_i(t)$ at each station $i$ is monitored.  
A sequence of nonzero 
occupation numbers at a given station constitutes a rain event, just as
in the radar observations \cite{peters}; the size of a rain event is
$s = \sum_t n_i(t)$ where the sum is over the set of consecutive
time steps for which $n_i (t) > 0$.  In case $n_i =0$, station $i$ is
said to experience a drought.  The durations of droughts and of
rain events are likewise monitored over a time interval $T$. 

As anticipated, the highly irregular velocity field
stretches and folds the tracer-bearing region \cite{aref84}.
At certain hyperbolic points the flow bifurcates.
Fig. 1 shows the particle positions at a time $\sim 1.3 \tau_C$,
in a system with 126 vortices ($\alpha = 3$).  The tracers are
widely scattered, but their distribution remains highly nonuniform.
At later times ($\sim 50\tau_C$), tracers are to be found  
throughout the system, but there are empty regions of various sizes
centered on vortices or vortex pairs.

The nonuniform tracer density leads to
scale-invariant rain and drought distributions.  Fig. 2 shows the 
number of rain events $N(s)$ as a function of size $s$, 
in a study with 126 vortices, $\alpha = 2$,
and $T \simeq \tau_c/2$ (the mean velocity $u \simeq 5$; statistics
are accumulated over a total of 200 realizations).  The
distribution follows a power law, $N(s) \sim s^{-0.95(1)}$ over nearly
five decades.  (Figures in parentheses denote uncertainties.) 
Fig. 3 shows the drought-duration distribution for
the same parameters; it follows a power law, with a decay exponent
of 1.14(1), over about three decades.  
The distribution
of rain {\it durations} (Fig. 3, inset) is more complicated,
decaying first as a power law (with an exponent of about 1), over
a decade or so, then attaining a roughly 
constant value, and finally decaying exponentially.

Note that for durations 
larger than $T/\Delta t = 10^4$, the drought 
distribution decays exponentially, as expected
for independent events.  The largest possible rain size
is $N_p T/\Delta t = 10^8$, corresponding to {\it all} of the tracers
sitting at a particular station for an entire simulation.  The largest 
rain event observed was $2.5 \times 10^6$, or 2.5\% the
maximum possible.  Given the small area of a station, this represents a
remarkable concentration of tracers for a long time,
reminiscent of blocking effects in the atmosphere, that may cause a
coherent stucture (possibly a vortex-dipole) 
to remain stationary over a long interval \cite{dritschel}.

Varying the vortex distribution and observation interval $T$,
the following trends emerge.  For $T/\tau_C$ in the range 0.1 - 2,
a power-law rain-size distribution, with an exponent in the
range of 0.93 - 1.02,
is observed over 4 - 5 1/2 decades.  The drought-duration distribution
decays with a somewhat larger exponent, 1.12 - 1.16, and follows
a power law over 3 - 4 decades.  Larger exponent values are
associated with higher values of $\alpha$; these yield somewhat
smaller ranges for the power laws.  Conversely, the largest power-law 
range, and smallest exponent values, are observed when all vortices
are of equal strength.
The is no significant difference between 
the distributions obtained initially
and those found after the vortices have had some time to 
evolve, suggesting that the equilibration process 
expected in two-dimensional turbulence \cite{kraichnan,aref}
is not important as regards rain and drought statistics \cite{rdunp}. 

Even systems with as few as ten vortices yield good power laws;
examples are shown in Fig. 4.
This indicates that chaotic advection
is the essential feature leading to scale invariance,
rather than well developed turbulence.
 
For larger values of $T/\tau_C$ the particles are more dispersed,
and the rain size and drought duration follow stretched-exponential
functions; an example is shown in Fig. 2 (inset) for
$T \simeq 200\tau_C$.
Even for large values of $T/\tau_C$ (up to 200 in the present study), 
the distributions decay more slowly than an exponential, 
showing that the tracer density is non-Poissonian.   

The simulation results may be summarized as
showing scale-invariant rain-size and drought-duration distributions
for intervals such that the tracers remain highly clustered.  
(Taking the scale of the original rain-formation region as $\sim $1 km,
the scale-invariant distributions correspond to rain
scattered over a region of $\sim 10 - 100$ km.  If we adopt a
typical wind velocity $u \sim 20$ km/h then the simulation time-step
$\Delta t$ is on the order of one second.)
Although the decay exponents
are somewhat smaller than those obtained from observational data
(1.36 and 1.42 for rain size and drought duration, resp.
\cite{peters}), the simulations also show the drought duration 
decaying more rapidly than that for rain event sizes.  For conditions under
which the rain is more thoroughly dispersed, simulations yield stretched-exponential
distributions.  While the latter have not been reported, it is well to
recall that the observational data come from a single station.  
Observations from other sites are needed confirm the generality of 
power laws and the possibility of other (non-scale-invariant) 
forms.

Clearly, the model employed in the present ``proof-of-principle" 
study contains a minimum of atmospheric
physics.  A three-dimensional description, 
allowing for stratification, convection, and vortex stretching would be 
desirable, as would inclusion of condensation, evaporation, and 
inertial effects \cite{benczik}.  
These improvements, all of which involve significant
computational complexity and expense, can be expected to alter detailed 
properties, and might well change the exponent values.  But since chaotic
advection is a
generic feature of atmospheric models, one should expect
scale-invariant distributions to appear.  This is supported by the
robustness of the power laws found in simulations,
to variations in number and strength of the vortices.
In this regard it is also interesting to note
that simulations of turbulent MHD processes reproduce power-law burst 
distributions for solar flares \cite{boffetta}, and that tracer 
patterns similar to those reported
here are also found in simulations of two-dimensional barotropic 
turbulence \cite{haller}.  Theoretical prediction of the rain and
drought distributions, and of the associated exponents, starting from
a model velocity field, remains as a formidable challenge.

In summary, I find that tracer distributions in 
two-dimensional flow, represented by a system of point
vortices, exhibit scale invariance during the early stage of the
dispersal process.
It therefore seems reasonable to attribute power-law 
rain and drought distributions
to turbulent atmospheric flow, and to develop more realistic models, 
to understand the observations in greater detail.

I thank Kim Christensen, Miguel A. Mu\~noz, Oscar N. Mesquita,
and Francisco F. Araujo Jr. for helpful discussions.
This work was supported by CNPq, and CAPES, Brazil.

\newpage
\noindent FIGURE CAPTIONS
\vspace{1em}

\noindent FIG. 1. Positions of 10$^4$ tracers at time $\simeq 1.3 \tau_C$,
in a system with 126 vortices, $\alpha = 3$.  The tracers are initially
distributed uniformly over the small square near the right edge.  Inset:
detail of a region of high tracer density, with two vortices.
\vspace{1em}

\noindent FIG. 2. Rain
size distribution in a study with 126 vortices, $\alpha = 2$,
and $T \simeq \tau_c/2$.  The straight line has slope -0.95.
Inset: stretched exponential rain size
distribution for $T \simeq 200\tau_C$, $\alpha = 4$. 
\vspace{1em}

\noindent FIG. 3. Drought-duration (main graph)
and rain-duration (inset)
distributions for the same parameters as Fig. 2.
The straight line has slope -1.14.
\vspace{1em}

\noindent FIG. 4. Rain-size (main graph) and drought-duration (inset)
distributions in a system of 10 vortices of equal strength,
$T \simeq 0.85 \tau_c$.  The stright lines have slopes of
-1.01 (rain size) and -1.13 (drought).
\vspace{1em}

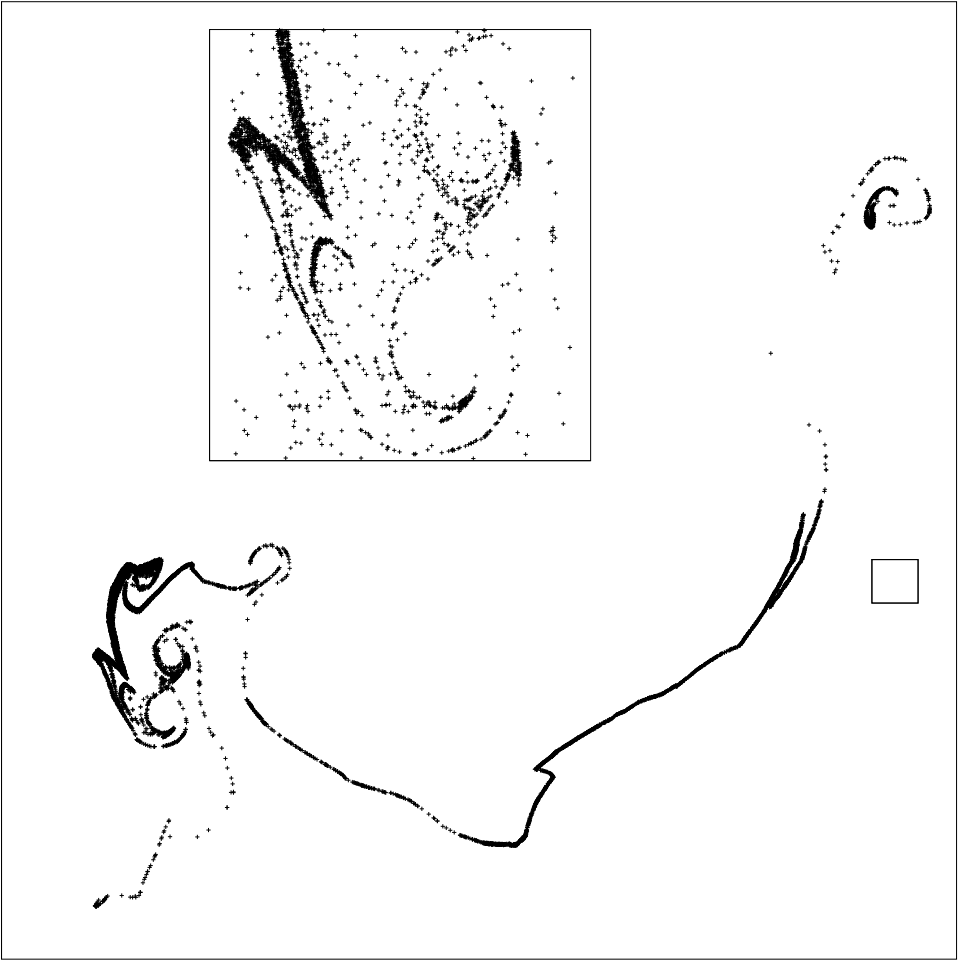
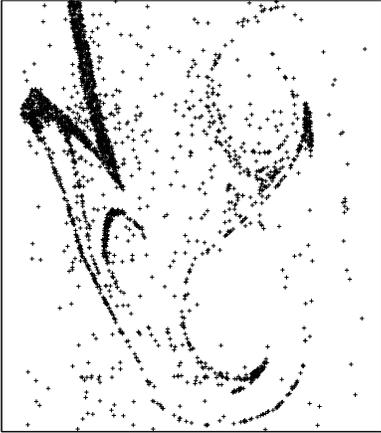
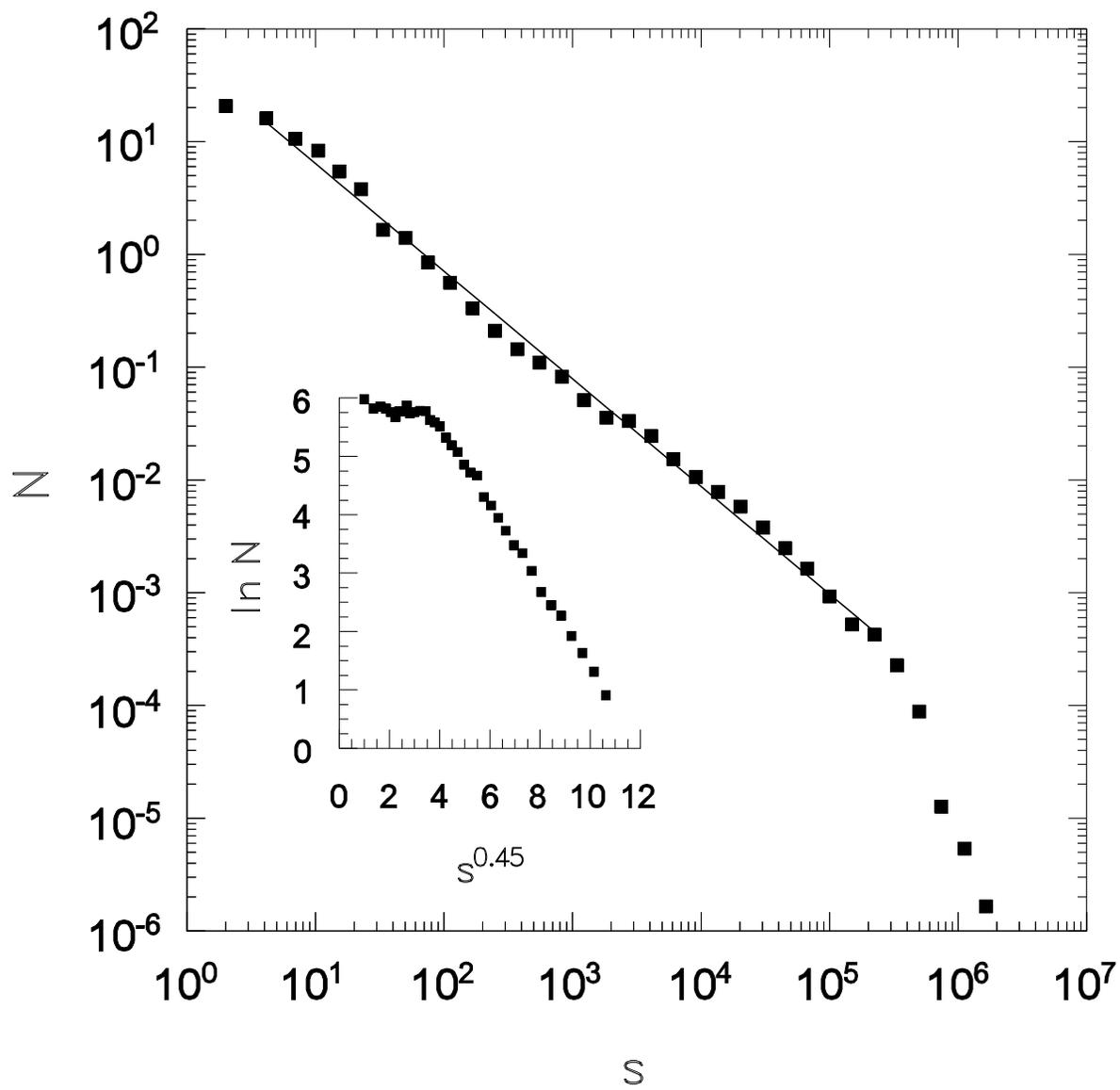
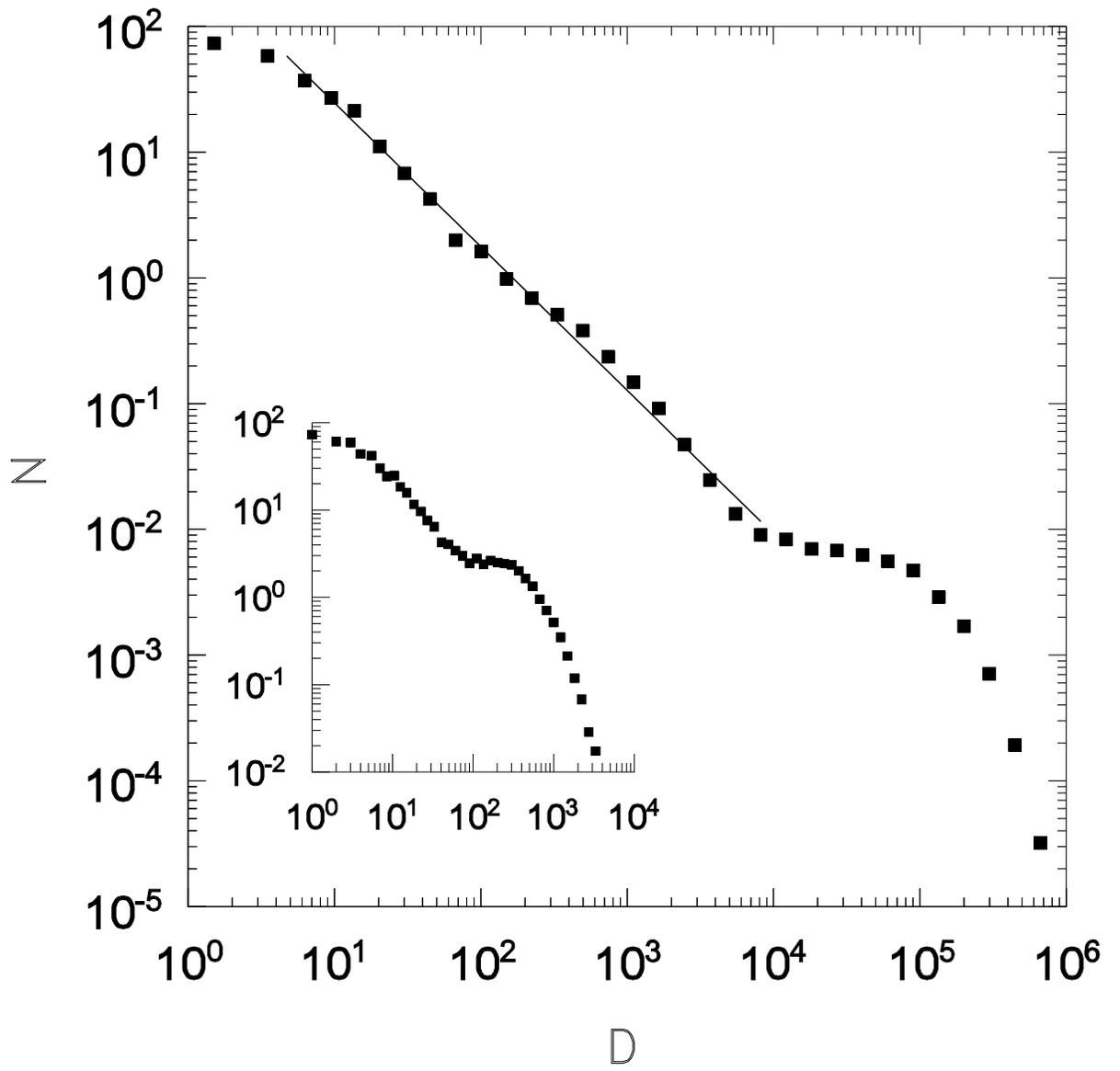
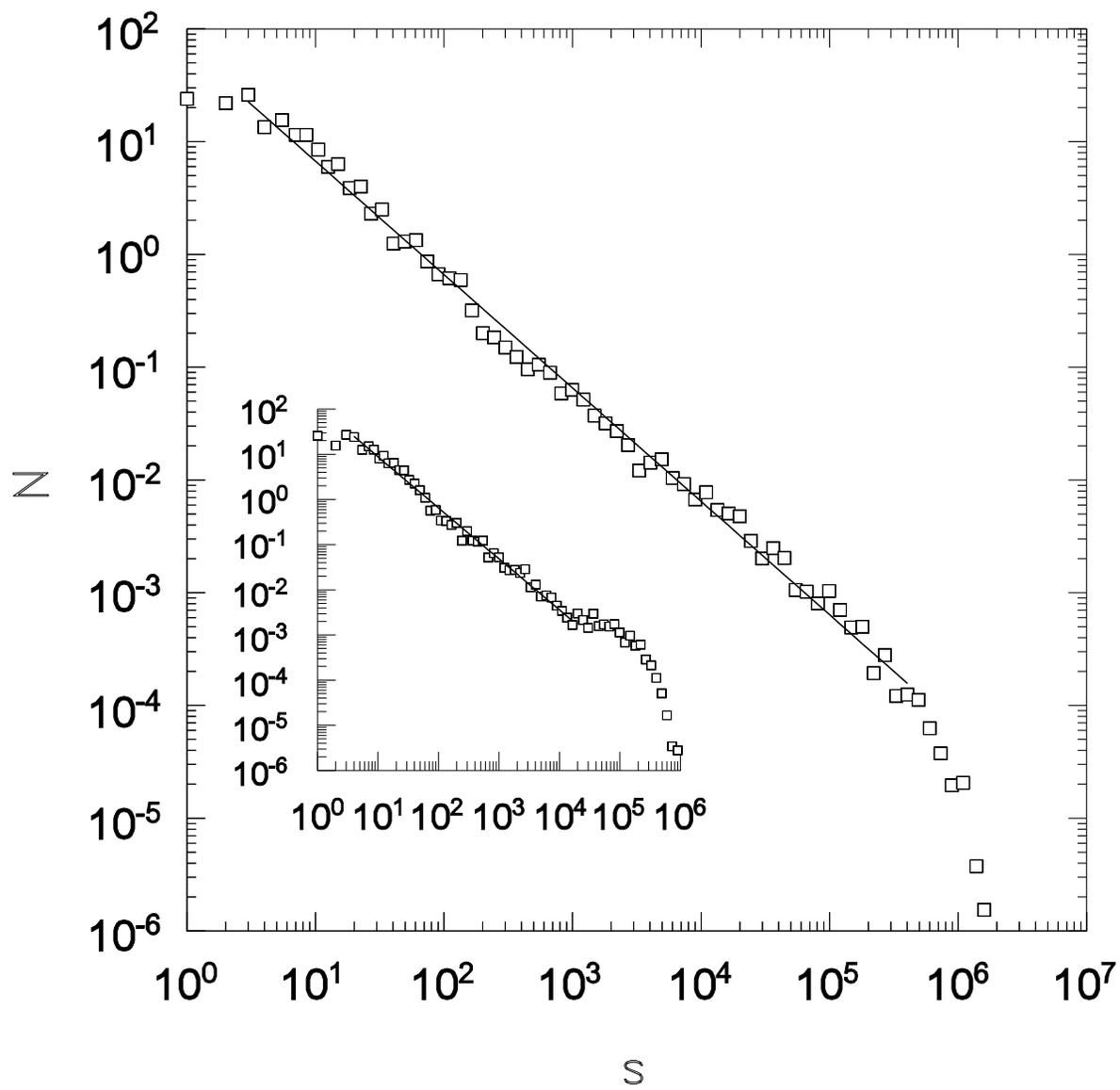
\end{document}